\documentclass{aa}

\usepackage{graphics,graphicx,subfigure}
\usepackage{natbib}
\usepackage{txfonts,bm}
\usepackage{color}
\begin{document}

\title{Fluffy dust forms icy planetesimals by static compression}


\author{Akimasa Kataoka\inst{1,2} \and Hidekazu Tanaka \inst{3} \and Satoshi Okuzumi \inst{4}\and Koji Wada \inst{5}}
\institute{
{Department of Astronomical Science, School of Physical Sciences, Graduate University for Advanced Studies (SOKENDAI), Mitaka, Tokyo 181-8588, Japan\\
\email{akimasa.kataoka@nao.ac.jp}
\and
National Astronomical Observatory of Japan, Mitaka, Tokyo 181-8588, Japan
\and
Institute of Low Temperature Science, Hokkaido University, Kita, Sapporo 060-0819, Japan}
\and
Department of Earth and Planetary Sciences, Tokyo Institute of Technology, Meguro, Tokyo, 152-8551, Japan
\and
Planetary Exploration Research Center, Chiba Institute of Technology, Narashino, Chiba, 275-0016, Japan
}
\abstract
 {
Several barriers have been proposed in planetesimal formation theory: bouncing, fragmentation, and radial drift problems.
Understanding the structure evolution of dust aggregates is a key in planetesimal formation.
Dust grains become fluffy by coagulation in protoplanetary disks.
However, once they are fluffy, they are not sufficiently compressed by collisional compression to form compact planetesimals.
}
{
We aim to reveal the pathway of dust structure evolution from dust grains to compact planetesimals.
}
{
Using the compressive strength formula, we analytically investigate how fluffy dust aggregates are compressed by static compression due to ram pressure of the disk gas and self gravity of the aggregates in protoplanetary disks.
}
{
We reveal the pathway of the porosity evolution from dust grains via fluffy aggregates to form planetesimals, circumventing the barriers in planetesimal formation.
The aggregates are compressed by the disk gas to a density of $10^{-3} {\rm g/cm^3}$ in coagulation, which is more compact than is the case with collisional compression.
Then, they are compressed more by self-gravity to $10^{-1} {\rm g/cm^3}$ when the radius is 10 km.
Although the gas compression decelerates the growth, the aggregates grow rapidly enough to avoid the radial drift barrier when the orbital radius is $\lesssim 6$ AU in a typical disk.
}
{
We propose a fluffy dust growth scenario from grains to planetesimals.
It enables icy planetesimal formation in a wide range beyond the snowline in protoplanetary disks.
This result proposes a concrete initial condition of planetesimals for the later stages of the planet formation.
}

\maketitle

\section{Introduction}
Planetesimals, the seeds of planets, are believed to form by coagulation of dust grains in protoplanetary disks.
How micron-sized dust grains grow to kilometer-sized planetesimals has been an unsolved problem in the complete planet formation theory; the intermediate-sized bodies are believed to be poorly sticky \citep{Zsom10}, easily disrupted by collisions \citep{Blum08}, or liable to fall quickly onto the central star \citep{Adachi76, Weidenschilling77}.

Several possibilities have been proposed to overcome these barriers \citep{Johansen07, Pinilla12a, Windmark12a, LambrechtsJohansen12, RosJohansen13, Garaud13}.
However, there has not yet been any coherent scenario explaining planetesimal formation from dust grains that avoids all of the barriers.

\begin{figure}[htbp]
 \begin{center}
  \includegraphics[width=80mm]{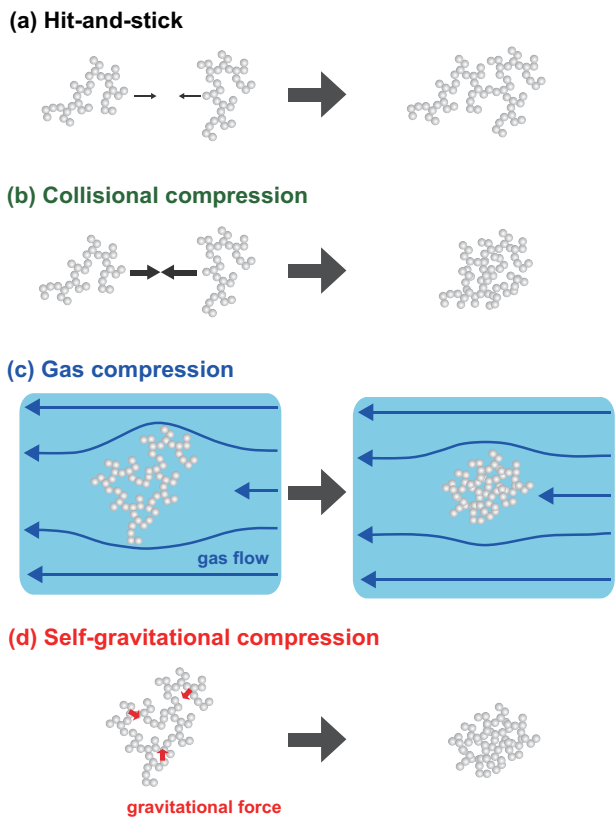}
 \end{center}
 \caption{
 Schematic drawing to illustrate dust growth via fluffy aggregates.
 (a) The dust aggregate hits another aggregate to be stick. 
 This reduces dust density and occurs in a very early stage of dust growth.
 (b) When the collisional speed is high enough to disrupt the dust aggregates, they are compressed.
 (c) Dust aggregates have a velocity difference against gas, and they feel the ram pressure by the gas.
 The ram pressure statically compresses the dust aggregates.
 (d) When the dust aggregates become so massive that they do not support their structure, they are compressed by their own self-gravity.
 }
 \label{fig:aggregate}
\end{figure}

The internal structure evolution is a key to understanding how dust coagulation forms planetesimals.
Figure \ref{fig:aggregate}(a) and (b) show the schematic diagram of the structure evolution previously considered.
Dust grains become porous aggregates composed of sub-micron monomer particles by coagulation in protoplanetary disks, as illustrated in Fig.\ref{fig:aggregate}(a) \citep{Smirnov90, Meakin91, Kempf99, BlumWurm00, KrauseBlum04, PaszunDominik06}.
When the dust aggregates become massive, they are gradually compacted or disrupted in dust-dust collisions because of the increase in the impact energy, as illustrated in Fig.\ref{fig:aggregate}(b)\citep{DominikTielens97, Wada07,Wada08,Wada09, Suyama08,Suyama12,PaszunDominik08, PaszunDominik09, Okuzumi12}.

Growth via fluffy aggregates has been proposed to be one possible scenario to overcome the barriers in \citet{Okuzumi12}.
They have shown that fluffy aggregates rapidly coagulate to avoid the radial drift problem.
On the other hand, although the aggregates are compressed by dust-dust collisions, their internal density remains $\rho\sim 10^{-5}{\rm g/cm^3}$ \citep{Suyama08, Okuzumi12}.
This is not consistent with the fact that planetesimals are believed to have $\rho\sim0.1{\rm g/cm^3}$ as well as comets, the remnants of planetesimals \citep{Ahearn11}.
Therefore other mechanisms to compress the fluffy aggregates are required.

In this paper, we introduce the static compression of aggregates due to ram pressure of the disk gas and self-gravity in protoplanetary disks, as illustrated in Fig.\ref{fig:aggregate}(c) and (d).
We use the compressive strength of porous aggregates numerically derived by \citet{Kataoka13} to obtain the porosity (equivalent to the internal density) of dust aggregates.
We show how much the dust aggregates are compressed by the disk gas and by self-gravitational compression in their growth.
Moreover, we investigate whether the growth is rapid enough to avoid the radial drift barrier by comparing the dust growth and radial drift timescale.

\section{Method}
The compressive strength of a highly porous dust aggregate, $P$, is given by \citep{Kataoka13}
\begin{equation}
P=\frac{E_{\rm roll}}{r_{0}^3}\left(\frac{\rho}{\rho_{0}}\right)^3,
\label{eq:eos}
\end{equation}
where $\rho$ is the mean internal density of the dust aggregate, $r_{0}$ the monomer radius, $\rho_{0}$ the material density, and $E_{\rm roll}$ the rolling energy, which is the energy for rolling a particle over a quarter of the circumference of another particle \citep{DominikTielens97,Wada07}.
In this paper, we adopt $\rho_{0}=1.0 {\rm ~g/cm^3}$, $r_{0}=0.1 {\rm ~\mu m}$, and $E_{\rm roll}=4.74 \times 10^{-9}$ erg , which correspond to icy particles.
$E_{\rm roll}$ is proportional to the critical displacement, which has an uncertainty from 2 ${\rm \AA}$ to 30 ${\rm \AA}$ \citep{DominikTielens97,Heim99}.
For later discussion, we note that the dust density is proportional to $E_{\rm roll}^{1/3}$ and thus the uncertainty little affects the resulting dust density.

When a dust aggregate feels a pressure that is higher than its compressive strength, the aggregate is quasi-statically compressed until its strength equals the pressure.
We define the dust internal density where the compressive strength equals a given pressure as an equilibrium density $\rho_{\rm eq}$.
Using Eq.(\ref{eq:eos}), we obtain $\rho_{\rm eq}$ as
\begin{equation}
\rho_{\rm eq}=\left(\frac{r_{0}^3}{E_{\rm roll}}P\right)^{1/3}\rho_{0}.
\end{equation}
We consider a source of the pressure to be ram pressure of the disk gas or self-gravity of the aggregate.

We obtain ram pressure of the disk gas as follows.
We consider a dust aggregate of mass $m$ and radius $r$, which is moving in the disk gas with velocity $v$ against the gas.
The pressure $P_{\rm gas}$ against the aggregate can be defined as the gas drag force divided by the geometrical cross section:  $P_{\rm gas}\equiv{F_{\rm drag}}/{A}$, where $F_{\rm drag}=mv/t_{\rm s}$, $A={\rm \pi} r^2$, and $t_{\rm s}$ is the stopping time of the aggregate.
While the pressure has both compressive and tensile components, we assume that the pressure is compressive.
Thus, we obtain the pressure as 
\begin{equation}
P_{\rm gas}=\frac{mv}{{\rm \pi}r^2}\frac{1}{t_{\rm s}}.
\end{equation}

\begin{figure}
 \begin{center}
  \includegraphics[width=80mm]{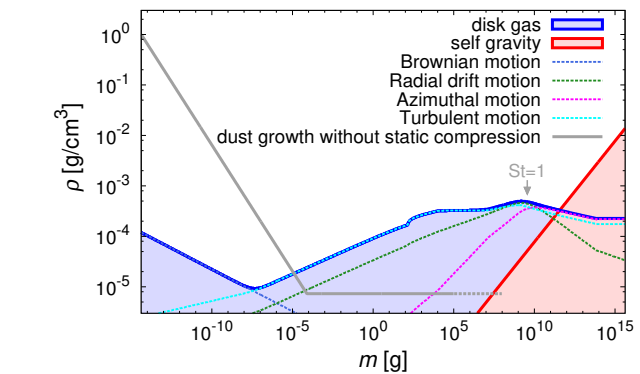}
 \end{center}
 \caption{
 Equilibrium dust density at 5AU in an MMSN disk.
 The blue thick solid line represents the equilibrium density of gas pressure, where the ram pressure of gas is equal to the compressive strength of the dust aggregate.
 The thin dotted lines represent the component of gas ram pressure, which is induced by the velocity difference between gas and dust, such as Brownian motion, radial drift motion, azimuthal motion, and turbulent motion.
 The red solid line represents the equilibrium density of self-gravity.
 The blue and red shaded region represents where the compressive strength of the dust aggregate is lower than the pressure of gas or self-gravity, so these aggregates are compressed until their density becomes the equilibrium density.
 We also plot the dust growth path without static compression \citep{Okuzumi12}.
 }
 \label{fig:rhoeq}
\end{figure}

The typical gas drag law is adopted to obtain $t_{\rm s}$ and $v$.
The gas drag law is the Epstein regime, when the dust radius is less than 4/9 times the mean free path of gas.
On the other hand, it is the Stokes regime if the Reynolds number is less than unity (see Eq.(4) in \citet{Okuzumi12}, for example).
When the Reynolds number exceeds unity, the gas drag law changes as a function of Reynolds number (see Eq.(8a) to Eq.(8c) in \citet{Weidenschilling77}).
The drag force is determined by the relative velocity of the gas and dust.
The relative velocity is induced by Brownian motion, radial drift, azimuthal drift, and turbulence.
We use the closed formula of the turbulence model \citep{OrmelCuzzi07} and assume the turbulent parameter $\alpha_{\rm D}=10^{-3}$, except for the strong turbulence case, where $\alpha_{\rm D}=10^{-2}$.

We assume the minimum mass solar nebula (MMSN), which was constructed based on our solar system \citep{Hayashi81}.
At a radial distance $R$ from the central star, the gas-surface density profile is $1700 {\rm ~g/cm^{2}} \times (R/1{\rm AU})^{-1.5}$ and the dust-to-gas mass ratio is 0.01.
The temperature profile adopted is $137 {\rm ~K} \times (R/1{\rm AU})^{-3/7}$, which corresponds to midplane temperature \citep{Chiang01}.
This is cooler than optically thin disk models to focus on the dust coagulation in the midplane.

We also calculate the self-gravitational pressure as follows.
Although the gravitational pressure has distribution in the aggregates, we simply assume a uniform pressure inside the aggregates.
We define the force on the dust aggregates as $F=Gm^{2}/r^{2}$, and the area $A=\pi r^{2}$.
Thus, the self-gravitational pressure is
\begin{equation}
P_{\rm grav}=\frac{Gm^2}{{\rm \pi}r^4}.
\end{equation}
We note that the equilibrium density of self-gravitational compression depends only on dust mass and internal density and not on the disk properties.

\section{Results}
First, we calculate the equilibrium density of dust aggregates in a wide range in mass, where their compressive strength is equal to the gas or self-gravitational pressure.
Figure \ref{fig:rhoeq} shows the equilibrium dust density against dust mass at 5 AU in the disk.
If the gas or self-gravitational pressure is higher than the compressive strength, the dust aggregate is compressed to achieve the equilibrium density because the strength is higher in denser dust aggregates.
In other words, the equilibrium density represents a lower limit of the dust density in the disk.
We also plot the collisional growth path without static compression \citep{Okuzumi12}; the dust aggregates initially grow with a fractal dimension of 2, and when the impact energy of dust-dust collisions reaches the rolling energy, the internal density becomes almost constant.
The evolutional track should trace the higher density of the growth with collisional compression and the equilibrium density of static compression.
Therefore, we conclude that dust growth is initially fractal, and then the gas compression becomes effective before collisional compression occurs at 5 AU in the disk.
\begin{figure*}
 \begin{center}
 	 \includegraphics[width=140mm]{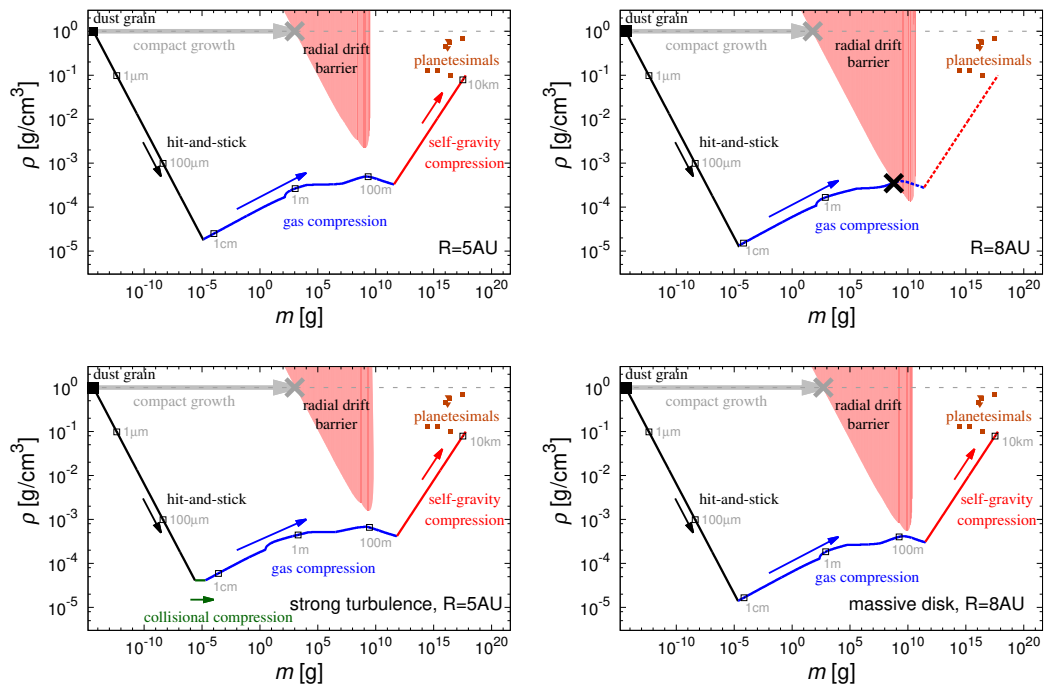}
 \end{center}
 \caption{
 Pathways in the planetesimal formation in the minimum mass solar nebula model.
 The gray line shows the constant density evolutional track, which corresponds to the compact growth.
 The black, green, blue, and red lines are the evolutional track through dust coagulation via fluffy aggregates.
 Each line represents different mechanisms of dust coagulation, which are hit-and-stick, collisional compression, gas compression, and self-gravity compression.
 The red shaded region represents where the radial drift timescale is less than the growth timescale, which is equivalent to radial-drift region.
 The brown squares indicate the properties of comets, and the triangles represent their upper limit.
 The radii of dust aggregates for 1 ${\rm \mu m}$, 1 cm, 1 m, 100 m, and 10 km are also written.
 (Top left): for 5 AU in orbital radius.
 (Top right): for 8 AU in orbital radius.
 The cross point represents where the dust falls onto the central star.
 (Bottom left): for 5 AU in strong turbulence model where $\alpha_{\rm D}=10^{-2}$.
 (Bottom right): for 8 AU in two times as massive as MMSN model.
 }
 \label{fig:path_rho}
\end{figure*}

Figure \ref{fig:path_rho} shows the pathways of the dust growth in mass-density space at 5 and 8 AU from the central star.
Here, we assume that the dust aggregates have no mass or volume distribution.
The first growth mode is hit-and-stick, where the fractal dimension is 2.
As dust aggregates become massive, the gas compression becomes effective.
Once gas compression occurs, the equilibrium density is higher in more massive aggregates until the Stokes number (i.e., the stopping time normalized by the orbital timescale) becomes unity; consequently, the aggregates always keep the equilibrium density in coagulation.
The gas compression remains effective until the dust aggregates grow so massive that self-gravitational compression is more effective than gas compression.
The self-gravitational compression is effective when the mass is $\sim 10^{11}$ g.
Once the gravitational compression is effective, the density increases with the mass of the power of 5/2.
However, the gas compression makes the dust density almost constant because of the constant velocity difference of the gas and dust of the head wind in azimuthal direction.
Therefore, the final stage is determined by self-gravitational compression.
We find that the dust aggregates should be compressed to a density of $0.1 {\rm ~g/cm^3}$ when their mass is $\sim 10^{18}$ g.

The density and mass of the final product are close to comets, which are considered to be the remnants of planetesimals.
We also plot the properties of several comets that are well known in the density and mass in Fig. \ref{fig:path_rho} \citep{Ahearn11}.
The comets have a mass of $\sim10^{16}$ g and an internal mass density of $\sim0.1{\rm ~g/cm^3}$.
There is a two-orders-of-magnitude discrepancy in mass between the final product of our calculation and the comets.
The mass of comets would be finally determined by collisional fragmentation or melting of planetesimals.

We estimate the dust growth timescale by coagulation and the radial drift timescale.
Assuming that aggregates have no mass or volume distribution, the growth time is defined as $m/\rho_{\rm d}\pi r^2 \Delta v$, where $\rho_{\rm d}$ is the spatial mass density of dust aggregates and  $\Delta v$ is the velocity difference between the dust aggregates, which is assumed be the root mean square of Brownian motion and turbulent motion (see Eq. (32) in \citet{Okuzumi12}).
For the velocity induced by turbulence, we use the velocity difference of dust and gas as dust-dust velocity for simplicity.
We include the dust sedimentation by considering the dust scaleheight \citep{Brauer08a}.
The drift timescale is defined as the orbital radius divided by the radial drift velocity.
We define the dust aggregates as where the drift timescale is less than 30 times the growth timescale drift to the central star \citep{Okuzumi12}.

We show that the revealed pathway still overcomes the radial drift problem.
Figure \ref{fig:path_rho} also illustrates the region where the dust aggregates radially drift inward before they grow.
Even when the dust aggregates become massive and their Stokes number is unity, the dust growth time is still shorter than the drift at 5 AU.
In the outer radius of the disk (e.g., at 8 AU), the dust aggregates drift inward before growth (top right of Fig. \ref{fig:path_rho}).
However, if the disk is two times as massive as MMSN, the dust aggregates successfully break the radial drift problem (bottom right of Fig. \ref{fig:path_rho}).
Therefore, the fluffy dust can grow to planetesimals more inside in protoplanetary disks.

Fluffy dust also breaks through the bouncing barrier.
Solid bodies have been shown to bounce in some situations \citep{Guttler10}.
However, both numerical simulations and experiments have shown that highly porous aggregates do not bounce when $\phi \lesssim 0.15$ \citep{Langkowski08, Wada11, SeizingerKley13, Kothe13}.
Thus, the bouncing is not a major difficulty for the growth of highly porous aggregates.

The fragmentation barrier is not a serious problem when considering icy particles.
The dust aggregates are not significantly fragmented and grow through collisions as long as their collision velocity is $< 60 {\rm~ m/s}$ \citep{Wada09}.
The collisional velocity becomes maximum when the Stokes number is unity.
It is the square root of $\alpha_{\rm D}$ times the sound speed of gas.
In the case of $\alpha_{\rm D} = 10^{-3}$, the maximum velocity is $\sim 17$ m/s at 5AU, so the dust aggregates can avoid significant disruption.
In the case of $\alpha_{\rm D} = 10^{-2}$, on the other hand, the velocity reaches $\sim 54$ m/s, comparable to the critical velocity.
Thus, we should carefully discuss the fragmentation in strong turbulent disks.
We note that the turbulent velocity is lower than the original MMSN model, where the temperature is determined by the balance between the stellar radiation and the reemission for each position.
This is because we focus on the midplane, where the temperature is lower; thus the velocity is also lower than the irradiated surface.

However, when considering silicate particles, it is difficult to break through the fragmentation barrier.
Inside the snowline, the ice is sublimated and the dominant material is silicate particles.
The critical velocity of fragmentation of silicate is $\sim 6 {\rm ~m/s}$ \citep{Wada09}.
Thus, the silicate planetesimal formation is still an open question.

In the final stage of coagulation, the runaway growth begins.
We estimate the dust mass where the dust-dust collision velocity induced by turbulence exceeds the escape velocity of the dust.
When the dust becomes as massive as $\sim10^{15}$ g, the runaway growth starts.
The dust internal density is still as small as $\sim 10^{-2}$, which means that the geometrical cross section is larger than the compact case.
This will make the runaway growth faster, but the whole scenario does not change, as shown in the $N$-body simulations \citep{KokuboIda96}. 

In conclusion, we revealed the pathway of the porosity evolution of dust aggregates to form planetesimals by introducing static compression.
We also showed that icy dust growth on the pathway avoids the bouncing, fragmentation, and radial drift barriers.
This scenario can provide a planetesimal distribution as a concrete initial condition of the later stages of planet formation.

\begin{acknowledgements}
A.K. is supported by the Research Fellowship from JSPS for Young Scientists (24$\cdot$2120).
\end{acknowledgements}

\bibliography{compression}
\clearpage

\end{document}